\documentclass[showpacs,amsmath,amssymb,twocolumn,pra,superscriptaddress]{revtex4-1}

\usepackage{amsfonts}
\usepackage{amssymb}
\usepackage{amscd}
\usepackage{amsmath}    % need for subequations
\usepackage{multirow}
\usepackage{graphicx}% Include figure files
\usepackage{dcolumn}% Align table columns on decimal point
\usepackage{bm}% bold math
\newcommand{\ket}[1]{\mbox{$\left| #1 \right\rangle$}}

\begin{document}
\preprint{APS/123-QED}
\title{Randomness generation based on spontaneous emissions of lasers}

\date{\today}% It is always \today, today,
\author{Hongyi Zhou}
%\email{yuanxiao12@mails.tsinghua.edu.cn}
\author{Xiao Yuan}
\author{Xiongfeng Ma}
\affiliation{Center for Quantum Information, Institute for Interdisciplinary Information Sciences, Tsinghua University, Beijing 100084, China}

%%%%%%%%%%%%%%%%%%%%%%%%%%%%%%%%%%%%%%%%%%
\begin{abstract}
Random number plays a key role in information science, especially in cryptography. Based on the probabilistic nature of quantum mechanics, quantum random number generators can produce genuine randomness. In particular, random numbers can be produced from laser phase fluctuations with a very high speed, typically in the Gbps regime. In this work, by developing a physical model, we investigate the origin of the randomness in quantum random number generators based on laser phase fluctuations. We show how the randomness essentially stems from spontaneous emissions. The laser phase fluctuation can be quantitatively evaluated from basic principles and also qualitatively explained by the Brownian motion model. After taking account of practical device precisions, we show that the randomness generation speed is limited by the finite resolution of detection devices. Our result also provides the optimal experiment design in order to achieve the maximum generation speed.
%Quantum random number generators can produce true randomness due to the probabilistic nature of quantum mechanics. One of the popular approaches of generating random number is based on phase fluctuations of a laser. In this work, we investigate the origin of randomness in such an approach. By studying a laser using a Brownian motion model, we examine the underneath mechanism of the phase fluctuation. We also take account of device imperfections, such as the finite resolution of data acquisition, to quantify the output randomness. To achieve more randomness, we optimize device parameters. Our result shows the fundamental limit on random number generation speed and suggests high-speed practical generator designs.
\end{abstract}

\vspace{2pc}
%\submitto{\NJP}
\maketitle

\section{Introduction}
Random numbers have vast applications in varieties of tasks, such as secure communication, numerical simulation and lottery. With the development of quantum information science, random numbers are also indispensable in many quantum information tasks. For example, in quantum key distribution (QKD) \cite{BB84}, random numbers are employed for the preparation of quantum states and the choice of measurement bases in order to guarantee its information-theoretical security.

Generally, there are two categories of physical random number generators (RNGs). One type is based on classical physics, whose randomness originates from the incomplete knowledge of the RNG system. For instance, classical RNGs can be based on chaotic behaviors of complicated systems such as semiconductor lasers \cite{Reidler2009}. Since a full characterization of the randomness generation process may enable an adversary to deterministically predict the outcomes, this type of RNGs cannot generate provable randomness. The other type with a mechanism of measuring a quantum state is called quantum random number generators (QRNGs). According to the basic principles of quantum mechanics, true random numbers can be generated, which means that the outcomes can never be predicted before the measurement.

\begin{figure}[bht]
  \centering
  \resizebox{6cm}{!}{\includegraphics{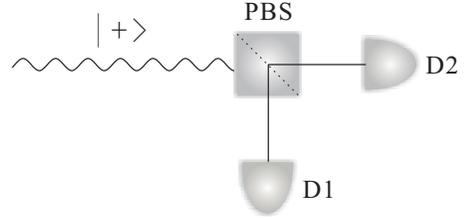}}
\caption{The simplest QRNG devices: a balanced polarization beam splitter and two single photon detectors $D_1$ and $D_2$.}\label{PBS}
\end{figure}

QRNGs have developed a lot in the past decade. The simplest QRNG is shown in Fig.~\ref{PBS}, which is designed by performing single photon detections.
Here, a single photon $\ket{+} = (\ket{H} + \ket{V})/\sqrt{2}$, which is a superposition of polarization $\ket{H}$ and $\ket{V}$, passes through a balanced polarization beam splitter and then it is detected by single photon detectors $D_1$ and $D_2$. The event of whether the detector $D_1$ ($D_2$) click or not for each photon is the desired output random number. Till now, many QRNGs are of this type, however, their generation speed is limited by the dead time of the detectors \cite{jennewein2000fast,stefanov2000optical}. Subsequently, some other types of QRNGs have been proposed. One of them is based on quantum non-locality, where randomness can be generated with very high security that even does not rely on the implementation \cite{pironio2010}. The price paying for such high security is that the generation speed is so low that the random numbers cannot satisfy most practical necessities. Another type measures the photon arrival time from a CW laser, and obtains random numbers from the timing measurement of single-photon detection relative to an external time reference \cite{wayne2010low,wahl2011ultrafast,nie2014practical}. Restricted by the count rate of the detectors and time resolution of the time-to-digital converters, the generation speed can not be very high. Recently, a new type of QRNGs measuring phase fluctuations from lasers \cite{Qi2010,jofre2011,symul2011,xu2012,abellan2014ultra} outperforms most of the existing ones in generation speed. Phase fluctuations is mainly caused by spontaneous emission of excited atoms of gain medium in a laser. As a direct effect of vacuum state fluctuations, spontaneous emission is an excellent resource for extracting true randomness. In experiment, such QRNG based on vacuum fluctuation is realized with a very high generation speed, up to 1Gbps \cite{xu2012,jofre2011}. It is also shown that the generation speed has the potential to be extended to 10 Gbps and even 100Gbps \cite{jofre2011}.

Such a high speed of the QRNG based on phase fluctuations leads to a natural question that whether there is a limit of the generation speed. To answer this question, in our work, we investigate the randomness generation mechanism by following a derivation  from first principles and use Brownian motion model to support it, which is a starting point to explore the limit of generation speed. We also build a simple model to express the generation speed as a function of some device parameters. We show that the generation speed is finitely limited when considering the precision or resolution of practical experiment instruments and give the optimal experiment design in order to achieve the maximum generation speed.

The rest of this paper is organized as follows. In Sec.~\ref{Sec:review}, we review the QRNG scheme and its underlying intuitive physical model. In Sec.~\ref{sec:variance}, we quantitatively derive the randomness output from first principles, which is also qualitatively explained  by comparing the similarity with the  Brownian motion model. In Sec.~\ref{Sec:optimization}, we show that the generation speed is finitely limited and give the optimal experiment design. Then we explain our result with device imperfection. We discuss another way quantifying randomness and conclude in Sec.~\ref{Sec:discussion}.

\section{QRNG based on vacuum fluctuations}\label{Sec:review}

\begin{figure}[htb]
  \centering
  \resizebox{8cm}{!}{\includegraphics{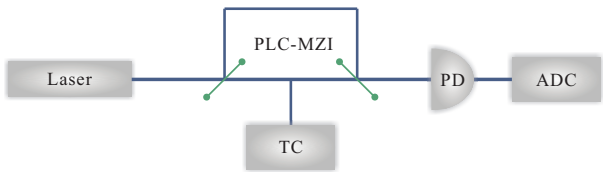}}
\caption{The experiment setup of the QRNG based on phase fluctuation.  (1) A 1.55$\mu m$ single mode cw distributed feedback (DFB) diode laser (ILX lightwave); (2) a compact planar lightwave circuit Mach-Zehnder interferometer (PLC-MZI) with a 500ps delay difference; (3) a temperature controller used to stabilize the delay difference of PLC-MZI; (4)a 5GHZ InGaAs photodetector; (5)an 8-bit analog-to-digital convertor (ADC).
}\label{DEVICE}
\end{figure}

In this section, we review a typical QRNG scheme based on measuring the vacuum fluctuations of a laser \cite{xu2012}.
As shown in  Fig.~\ref{DEVICE}, a laser beam passes through a planar lightwave circuit Mach-Zehnder interferometer (PLC-MZI). Due to the spontaneous emission, there will be a phase fluctuation in the laser field compared to a plane wave function. The phase fluctuation is transformed into intensity fluctuation by the interferometer and then detected by the photodetector (PD). The output voltage from PD will be sampled and digitized by an 8-bit analog-to-digital convertor (ADC), which will produce true random numbers.

To get the expression of the output randomness, let us first suppose that the lasing field  can be expressed by
\begin{equation}\label{Eq:ESPhi}
  E(t)=\sqrt{S} \exp[i(\phi_0(t) + \phi'(t))],
\end{equation}
where $\phi_0(t)$ represents the phase of a plane wave, $\phi'(t)$ represents the phase fluctuations due to spontaneous emission as well as some other factors, and $S$ represents the photon number intensity of $E$.
Suppose the time delay between the two arms of the PLC-MZI is $\tau_l$, the measurement output voltage $V(t)$, which is proportional to the interference probability, can be  expressed by
\begin{equation}\label{00}
V(t) \propto 2E^*(t)E(t+\tau_l)\sin(\Delta \phi(t)) \propto P \Delta \phi'(t),
\end{equation}
where $P$ is the laser output power and $\Delta \phi'(t)=\phi'(t)-\phi'(t+\tau_l)$ is the total phase fluctuations.

The variance of the phase fluctuations is modeled by
 \begin{equation}\label{Eq:variancetheta}
\langle \Delta \phi'(t)^2 \rangle =\frac{Q}{P}+C,
\end{equation}
where ${Q}/{P}$ and $C$ stand for the contributions from spontaneous emission and classical mechanisms, respectively. In experiment, in order to detect such a variance, we transform it into the variance of the output voltage by multiplying Eq.~\eqref{Eq:variancetheta} by $P^2$
\begin{equation}\label{Eq:varianceV}
\langle V^2 \rangle=AQP+ACP^2+F,
\end{equation}
where $A$ is a constant coefficient and $F$ is the background noise. The value of $AQ$, $AC$ and $F$ can be obtained by fitting  experiment data of $P$ and $\langle V^2 \rangle$ according to Eq.~\eqref{Eq:varianceV}. Then, the laser power $P$ is chosen to maximize the quantum signal to classical noise ratio $\gamma$
\begin{equation}\label{7}
\gamma=\frac{AQP}{ACP^2+F}.
\end{equation}

\begin{figure}[htbp]
\centering
\resizebox{8cm}{!}{\includegraphics{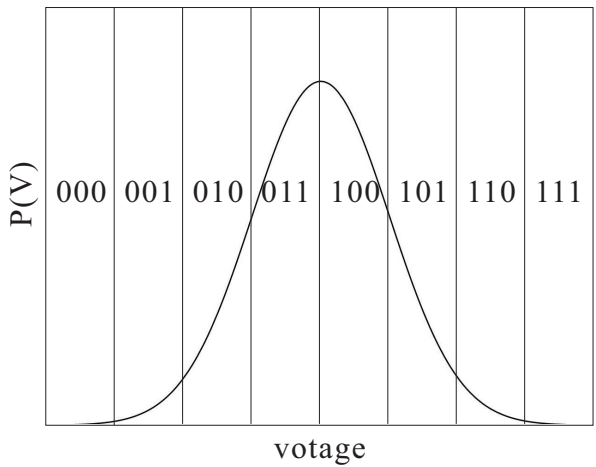}}
\caption{Probability distribution of the output voltage. The $x$-axis stands for the output voltage and the $y$-axis stands for the probability density of the voltage. The value range of the output voltage will be divided to 8 intervals. If the voltage is measured to be in some interval, a 3-bit sequence of random numbers will be generated accordingly. However, due to the classical noises, we use $\langle V^2_{Q} \rangle$ instead of $\langle V^2\rangle$ to calculate the maximum probability.}
\label{ADC}
\end{figure}

The process to obtain random numbers is shown in Fig.~\ref{ADC}, where we use a 3-bit ADC instead of an 8-bit one for simplicity. The output voltage is measured and discretized by dividing its value range into 8 intervals. Each voltage interval corresponds to a 3-bit random number sequence. An 8-bit ADC works similarly in practice.

Finally, the method to quantify the random number generation speed is as follows. First it is necessary to quantify the output randomness per sample with the min-entropy function,
\begin{equation}\label{Eq:min-entropy}
  H_{\infty}(X)=-\log_2\left(\mathrm{max}_{x \in \{ 0,1 \}^n}\Pr[X=x]\right),
\end{equation}
where $X$ stands for the $n$ random numbers obtained .
When assuming that the voltage follows a Guassian distribution, as the true randomness comes only from the quantum contribution $AQP$ in Eq.~\eqref{Eq:varianceV}, the maximum probability $\Pr[X=x]$ can be calculated from the quantum variance $\langle V^2_{Q} \rangle$,
\begin{equation}\label{Eq:vq^2}
\begin{aligned}
\langle V^2_{Q} \rangle=\frac{\gamma}{\gamma+1} \langle V^2 \rangle.
\end{aligned}
\end{equation}
The maximum probability $\Pr[X=x]$ equals the area of the biggest part under the Guassian function curve $G(0,\sqrt{\langle V^2_{Q} \rangle})$. Then we can calculate the output randomness per sample $R_0$ and the generation speed is the product of $R_0$ and the sampling rate.

Therefore we can obtain the final randomness generation rate by making the following assumptions,
\begin{enumerate}
  \item The phase difference $\langle \Delta \phi'(t)^2 \rangle$ is assumed to satisfy Eq.~\eqref{Eq:variancetheta}.
  \item The output voltage $V$ follows a Guassian distribution.
  \item The random numbers generated from different samples are mutually independent.
\end{enumerate}

In our work, we start from first principles to theoretically derive these assumptions. In the original experiment \cite{xu2012}, the final QRNG speed is given by a product of the sampling frequency and the randomness output in each sample. It seems that the generation speed can infinitely increase with infinite sampling speed. While, we show a counter result. For given experiment setups, the generation speed is finite and maximized with a finite sampling speed.

\section{Improvement of previous QRNG physical model }\label{sec:variance}
In this section, we will start with randomness origin in laser source and then derive the assumptions 1 and 2 with basic physical models from first principles. In assumption 1, the quantum contribution of phase fluctuations $Q$ in Eq.~\eqref{Eq:variancetheta} is a constant for given experimental parameters. Our result give an accurate expression of $Q$ and show a linear relationship between $Q$ and $\tau_l$. For assumption 2, we prove that the output voltage V follows a Guassian distribution indeed. Moreover, we will use a Brownian motion model to describe the behavior of the random phase and explain our result from an intuitive perspective.

\subsection{Randomness origin in laser source}

Let us first introduce the basics of lasers and the origin of the randomness from a laser source. The structure of a laser source mainly consists of an optical resonator and gain medium.
The gain medium can be regarded as ensembles of charged particles with two level energy levels denoted by a ground state $\ket{g}$ and an excited state $\ket{e}$. When a laser source is below the threshold, the population of the particles follows a Boltzmann distribution. If a population inversion is achieved due to some external factors, the laser source is above the threshold and begins to generate laser.  Considering the interaction between atomic electron and electromagnetic field, a population inversion will lead to two kinds of photon emissions: stimulated emission and spontaneous emission. The former is of perfect monochromaticity with energy given by
\begin{equation}\label{Eq:nu}
  h\nu = E_e - E_g,
\end{equation}
where $E_e$ and $E_g$ are the energies of the excited and ground level, respectively.
And the latter is the origin of randomness.

The output state of spontaneous emission can be given by
\begin{equation}
\left| \Psi(t)\right \rangle =a(t)e^{-i \omega_0 t}\left| e; 0\right \rangle + \sum_{k,s} b_{ks}(t) e^{-i \omega_k t}\left|g; 1 \right \rangle,
\end{equation}
according to the Weisskopf-Wigner theory \cite{scully1997quantum}. Here $\ket{0}$ and $\ket{1}$ denote the vacuum and single photon state respectively, $a(t)$ and $b_{ks}(t)$ are the probability amplitude, $k$ stands for the wave vector, $\omega_k=ck$ is the frequency of the photon, and $s$ refers to the spin of the charged particle. As $\omega_k$ generally differs from $\nu$ defined in Eq.~\eqref{Eq:nu}, measuring the phase of the emitted photons produces random numbers with a similar mechanism as the single photon QRNG scheme illustrated in Fig.~\ref{PBS}.

\subsection{Variance of the quantum phase fluctuations}

Now, we derive the variance expression in Eq.~\eqref{Eq:variancetheta} from first principles. To derive the dynamic evolution of the laser field, let us consider the  laser field changes during an infinitesimal time interval $\delta t$. Taking account of the stimulated and spontaneous emission, the laser field change can be given by
\begin{equation}\label{Eq:laserfiledchange}
E(t+\delta t) - E(t) = \left[ f\left(E(t)\right) + E_{\mathrm{sp}}(t)\right]\delta t,
\end{equation}
where $E_{\mathrm{sp}}(t)$ corresponds to the laser field from spontaneous emission and $f\left(E(t)\right) $ corresponds to the contributions of stimulated emission and any other factors that may affect the laser field. Thus, the dynamic evolution of the laser field can be given by \cite{petermann1991}
\begin{equation}\label{Eq:laserdynamic}
\frac{\mathrm{d}E(t)}{\mathrm{d}t}=f\left(E(t)\right) +E_\mathrm{sp}(t).
\end{equation}

For the laser field $E$ expressed in Eq.~\eqref{Eq:ESPhi}, the total phase $\phi(t)$ can be rewritten in detail
\begin{equation}
\phi(t)= \phi_0(t) + \phi_1(t) + \phi_{\mathrm{sp}}(t),
\end{equation}
where $\phi_1(t)$ and $\phi_{\mathrm{sp}}(t)$ represent the classical and quantum phase noise respectively.
The total phase dynamically evolves according to
\begin{equation}
\begin{aligned}
\frac{\mathrm{d}\phi(t)}{\mathrm{d}t}&=\frac{1}{S(t)}\mathrm{Im}(E^*(t)\frac{\mathrm{d}E(t)}{\mathrm{d}t})\\
&=\frac{\mathrm{Im}\left[E^*(t)f\left(E(t)\right) \right]}{S(t)}+\frac{\mathrm{Im}\left[E^*(t)E_\mathrm{sp}(t)\right]}{S(t)}\\
&=\frac{\mathrm{Im}\left[e^{-i\phi(t)}f\left(E(t)\right) \right]}{\sqrt{S(t)}}+\frac{\mathrm{Im}\left[e^{-i\phi(t)}E_\mathrm{sp}(t)\right]}{\sqrt{S(t)}}.
\end{aligned}
\end{equation}
The evolution of the phase $\phi_{\mathrm{sp}}(t)$ is thus given by
\begin{equation}
\frac{\mathrm{d} \phi_{\mathrm{sp}}(t)}{\mathrm{d}t}=\frac{\mathrm{Im}\left[e^{-i\phi(t)}E_\mathrm{sp}(t)\right]}{\sqrt{S(t)}}=F_{\mathrm{sp}}(t).
\end{equation}
Due to the Weisskopf-Wigner theory, the emitted field $E_\mathrm{sp}$ by spontaneous emission is a white noise. Thus, we can regard $F_{\mathrm{sp}}(t)$ as i.i.d. for different time.

In the experiment setup in the last section, the time delay between the two arms in the
PLC-MZI is $\tau_l$. Then the phase difference $\Delta \phi_{\mathrm{sp}}$ induced by quantum spontaneous emission is
\begin{equation}\label{Eq:Deltaphi}
\Delta \phi_{\mathrm{sp}}= \phi_{\mathrm{sp}}(t_0+\tau_l)- \phi_{\mathrm{sp}}(t_0)=\int_{t_0}^{t_0+\tau_l} F_{\mathrm{sp}}(t) \mathrm{d}t.
\end{equation}
The variance of the quantum phase fluctuation can be written as
\begin{equation}\label{Eq:variance}
\begin{aligned}
\langle \Delta \phi_{\mathrm{sp}}^2 \rangle & =\left\langle \left(\int_{t_0}^{t_0+\tau_l}F_{\mathrm{sp}}(t) \mathrm{d}t\right)^2 \right\rangle \\ &=\left\langle  \left(\int_{t_0}^{t_0+\tau_l} F_{\mathrm{sp}}(t)\mathrm{d}t\right) \left(\int_{t_0}^{t_0+\tau_l} F_{\mathrm{sp}}(t')\mathrm{d}t'\right)  \right\rangle \\ &= \int_{t_0}^{t_0+\tau_l} \left\langle F_{\mathrm{sp}}(t) F_{\mathrm{sp}}(t')\right\rangle \mathrm{d}t \mathrm{d}t' \\ &=\frac{R_\mathrm{sp}}{2\langle S\rangle}\int_{t_0}^{t_0+\tau_l} \delta(t-t')\mathrm{d}t \mathrm{d}t'  \\ &=\frac{R_\mathrm{sp}\tau_l}{2\langle S\rangle}.
\end{aligned}
\end{equation}
Here, we make use of the autocorrelation of $F_{\mathrm{sp}}(t)$
\begin{equation}\label{Eq:F}
\langle F_{\mathrm{sp}}(t)F_{\mathrm{sp}}^*(t-\tau_l)\rangle=\frac{R_\mathrm{sp}\delta(\tau_l)}{2\langle S\rangle},
\end{equation}
which is derived from the aforementioned white noise assumption with an autocorrelation function
\begin{equation}
\langle E_{\mathrm{sp}}(t)E_{\mathrm{sp}}^*(t-\tau_l) \rangle =R_\mathrm{sp} \delta (\tau_l).
\end{equation}
We refer to Ref.~\cite{petermann1991} for detail derivation of Eq.~\eqref{Eq:F}.

So we can draw a conclusion with the deduction above that the variance of the quantum phase fluctuations $\langle \Delta \phi^2 \rangle \propto \tau_l$. Moreover, the same result can be interpreted physically with a Brownian motion model.

Brownian motion is a physical phenomenon which describes the random motion behavior of tiny particles suspended in a fluid. Due to the collision of the molecules in the fluid, the tiny particles will move randomly. Einstein's theory about the Brownian motion shows that the variance of the displacement of the tiny particle is linearly proportional to the elapsed time, which is very similar to Eq.~\eqref{Eq:phivariance}.

\begin{figure}[htbp]
\centering
\resizebox{6cm}{!}{\includegraphics{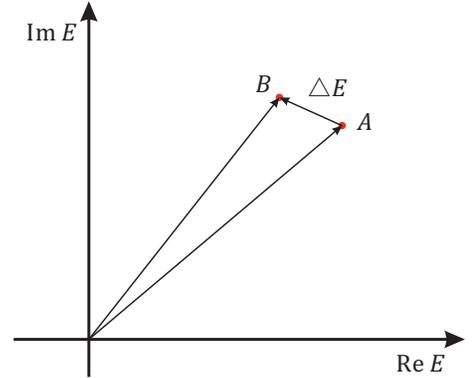}}
\caption{The laser field changes $\Delta E$ for the increment of a single photon. The length of $\Delta E$ is equal to 1 and the phase of $\Delta E$ is completely random. It has the same physical picture as a step of a two dimensional random walk from $A$ to $B$, which has been proved to be a Brownian motion. }
\label{SP}
\end{figure}

Considering the complex laser field as a vector in two-dimensional coordinates, when a photon is added to the field by spontaneous emission, as shown in Fig.~\ref{SP}, there will be a small shift both in phase and amplitude  \cite{yariv2006}. Due to the property of spontaneous emission, each increment caused by a photon emitted is completely random in direction and with a constant step length.  Thus, the lasing field vector can be regarded as a two dimensional random walk, which can be proved to be a Brownian motion process \cite{ross2014introduction}. Then we can draw the same conclusion that the variance of the quantum phase fluctuations $\langle \Delta \phi^2 \rangle \propto \tau$ with the characteristics of Brownian motion process.

Generally, when a laser source is above the threshold, its output power $P$ is proportional to the photon number intensity $S$ \cite{petermann1991}
\begin{equation}
P=C{\langle S\rangle},
\end{equation}
where $C$ is a constant determined by intrinsic parameters of a laser diode. Thus, we derive the first assumption
\begin{equation}\label{Eq:phivariance}
  \langle \Delta \phi^2_{\mathrm{sp}} \rangle = \frac{CR_\mathrm{sp}\tau_l}{2 P}.
\end{equation}
Compared with Eq.~\eqref{Eq:variancetheta}, we can easily get
\begin{equation}
Q=\frac{CR_\mathrm{sp}\tau_l}{2}
\end{equation}
As for assumption 2, to see why the voltage $V$ follows a Gaussian distribution, we only need to prove that the phase difference $\Delta \phi_{\mathrm{sp}}$ follows the same distribution according to Eq.~\eqref{00} and  Eq.~\eqref{Eq:vq^2}. From the definition in Eq.~\eqref{Eq:Deltaphi}, we can think that $\Delta \phi_{\mathrm{sp}}$ is the sum of $N$ identically independent distributed (i.i.d.) variables, where each one is given by
\begin{equation}\label{Eq:i.i.d}
\int_{t}^{t+\tau_l/N} F_{\mathrm{sp}}(t') \mathrm{d}t'.
\end{equation}
Due to the central limit theorem, we can easily conclude that $\Delta \phi$ follows a Gaussian distribution and its variance is given in Eq.~\eqref{Eq:variance}.

\section{Maximize output randomness and generation speed}\label{Sec:optimization}
In previous experiments, the random number generation speed is the product of the output randomness per sample and the sampling rate. Even with perfect detectors, the QRNG speed is limited by the sampling rate \cite{MXF:2013:RandomPost}. That is, the output randomness per sample is finite. On the other hand, it seems that the generation speed will infinitely increase with infinite sampling rate. In this section, we calculate the randomness output for given experiment setups. We show that the output randomness and random number generation speed are maximized with finite sampling rate when other experiment parameters are fixed.

\begin{figure}[htbp]
\centering
\resizebox{8cm}{!}{\includegraphics{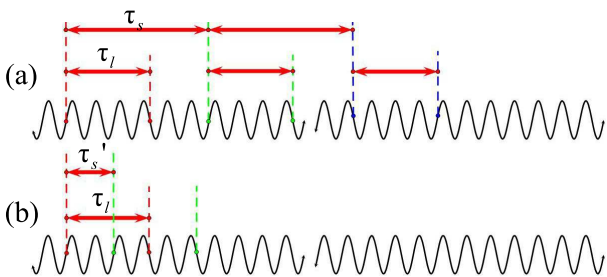}}
\caption{(a) The output of the PLC-MZI in each sample can be regarded as a interference from points A and B in the same beam with time difference $\tau_l$. (b) When the sampling rate increases, there will be an overlap between different samples, and the quantum randomness $\Delta \phi_{\mathrm{sp}}$ is not mutually independent for different samples.}
\label{OVERLAP}
\end{figure}

In our analysis, the time delay $\tau_l$ and the sampling time interval $\tau_s$ are free variables. As the PLC-MZI interferes two laser beams whose emission time difference is $\tau_l$, we can equivalently regard it as that the interference comes from two points in the same beam with time difference $\tau_l$. As shown in  Fig.~\ref{OVERLAP}~(a), we can think the point $A$ and $B$  interfere for example.

When  $\tau_l \leq \tau_s$, the quantum randomness $\Delta \phi_{\mathrm{sp}}$ in Eq.~\eqref{Eq:Deltaphi} are independent between different samples. This is because the integral of  $F_{\mathrm{sp}}(t)$, which is i.i.d., does not overlap between different samples. We thus show the assumption 3 is correct for $\tau_l \leq \tau_s$. It is straightforward to see that, the output randomness will increase with increasing sampling speed as long as the condition $\tau_l \leq \tau_s$ holds.

On the other hand, as shown in Fig.~\ref{OVERLAP}~(b), when $\tau_l > \tau_s$, the quantum randomness $\Delta \phi_{\mathrm{sp}}$ are not independent for different samples. The random output $\Delta \phi_{\mathrm{sp}}$ for successive samples have overlapped integrals, thus are dependent on each other. In this case, we disprove the assumption 3 for $\tau_l > \tau_s$. When the sampling rate is high enough, we can think $\tau_l $ is an integer multiple of $\tau_s$. Then as shown in Appendix, this situation can be regarded as that the random numbers are generated with a smaller $\tau_l'$ that $\tau_l'=\tau_s$, and we can equivalently quantify the output randomness with the parameters $\tau_l'$ and $\tau_s$.

\begin{figure}[htbp]
\centering
\resizebox{8cm}{!}{\includegraphics{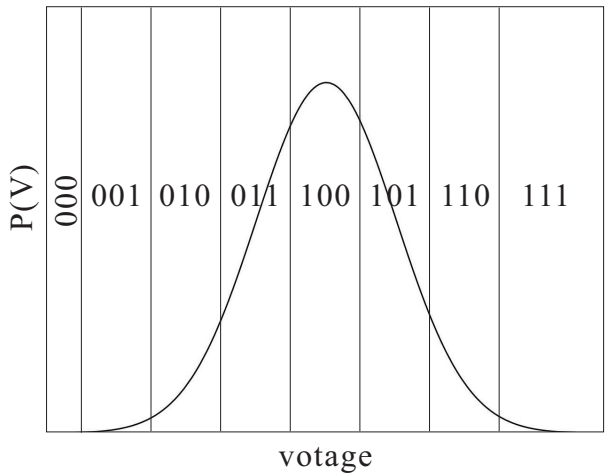}}
\caption{A shift of the voltage partition caused by classical noise and potential adversaries will always reduce the output randomness per sample, and the randomness has a minimum with the partition here, which represents the worst case.}
\label{SHIFT}
\end{figure}

Thus, when considering the maximum production of randomness, we can always focus on the case of $\tau_l = \tau_s$. Then we just need to optimize $\tau_l$ to achieve the most output randomness.
First, the coherence time $\tau_c$ can be defined by that the variance of $\phi_{\mathrm{sp}}$ changes $(2\pi)^2$,
\begin{equation}\label{6}
\langle \Delta \phi^2 \rangle= \frac{CR_\mathrm{sp}\tau_c}{2 P}  =(2\pi)^2.
\end{equation}
In this case, we can rewrite Eq.~\eqref{Eq:phivariance} by
\begin{equation}\label{Eq:phivariance2}
\langle \Delta \phi^2 \rangle=4\pi^2\frac{\tau_l}{\tau_c},
\end{equation}
and the variance of the output voltage is given by
\begin{equation}\label{Eq:VQ^2vstau_l}
\langle V^2_{Q} \rangle= AP^2\langle \Delta \phi^2 \rangle=4AP^2\pi^2\frac{\tau_l}{\tau_c}.
\end{equation}
The output voltage follows a Gaussian distribution, and the output random number is from the voltage in a certain interval. In the practical scenario where there might exist classical noises and potential adversaries, the minimum randomness for each sampling, as shown in Fig.~\ref{SHIFT}, can be given by
\begin{equation}\label{Eq:randomnessineachsample}
\begin{aligned}
R_0&=-\log_2{\left[\Phi\left(\frac{a}{2\sqrt{\langle V^2_{Q} \rangle}}\right)-\Phi\left(-\frac{a}{2\sqrt{\langle V^2_{Q} \rangle}}\right)\right]},\\
&=-\log_2{\left[2\Phi\left(\frac{\lambda}{\sqrt{\tau_l}}\right)-1\right]},
\end{aligned}
\end{equation}
where $\Phi(x)$ is the cumulative distribution function of a standard  Gaussian Distribution, $a$ is the length of a voltage interval, and $\lambda$ is a constant determined by experiment parameters according to
\begin{equation}
  \lambda = \frac{a}{4\pi P}\sqrt{\frac{\tau_c}{A}}.
\end{equation}

\begin{figure}[htbp]
\centering
\resizebox{8cm}{!}{\includegraphics{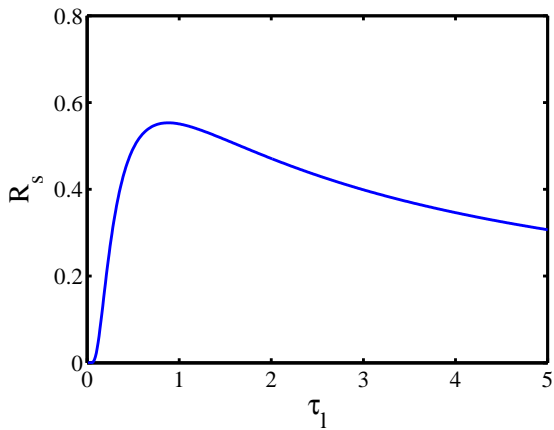}}
\caption{The unit of the random number generation speed $R_s$ and the time delay difference $\tau_l$ are $\mathrm{bit/s}$ and $\mathrm{s}$ respectively. To merely show their relationship, we set the constant $\lambda$ to be $1\mathrm{\sqrt{s}}$ for simplicity. Then we can see that $R_s$ has a maximum at $\tau_l=0.88\mathrm{s}$. In experiment, the actual value of the optimized $\tau_l$ can be calculated similarly with an accurate $\lambda$.}
\label{tau_l}
\end{figure}

In a fixed sample time $T$, the total output randomness is
\begin{equation}\label{Eq:Rtot}
\begin{aligned}
R_{\mathrm{tot}}&=\frac{T}{\tau_s}R_0,\\
&=-\frac{T}{\tau_l}\log_2{\left[2\Phi\left(\frac{\lambda}{\sqrt{\tau_l}}\right)-1\right]},
\end{aligned}
\end{equation}
where we make use of the relation $\tau_l = \tau_s$. Moreover, if we multiply Eq.~\eqref{Eq:Rtot} by $1/T$, the result stands for the random number generation speed $R_s$,
\begin{equation}
R_s=-\frac{1}{\tau_l}\log_2{\left[2\Phi\left(\frac{\lambda}{\sqrt{\tau_l}}\right)-1\right]}.
\end{equation}
Here, we show the relationship between $R_s$ and $\tau_l$ in Fig.~\ref{tau_l}. We can see that the generation speed $R_s$ has a maximum.

The existence of maximum generation speed can be intuitively explained, which is mainly owing to the finite resolution of the ADC. In Eq.~\eqref{Eq:phivariance2}, we can see that the variance of the detected random number is linearly proportional to the sampling time interval when $\tau_s = \tau_l$. With increasing sampling speed, the sampling time interval $\tau_s$ will decrease. When $\tau_s$ is small enough, the detected random numbers will lie almost always in the same interval of the ADC. In other words, we will not detect the quantum noise. In this case, the output randomness in each sample is almost zero as can be inferred from Eq.~\eqref{Eq:randomnessineachsample}. Therefore, considering the resolution of ADC, the output randomness will be upper bounded with a finite value. Additionally, in experiment, if a high resolution ADC is not available, we will add a high gain amplifier after the photo-detector, which is equivalently to increase parameter $A$ in Eq.~\eqref{Eq:varianceV}. However, combining Eq.~\eqref{Eq:varianceV} and Eq.~\eqref{Eq:VQ^2vstau_l} we can see that the additional classical noise introduced will increase with sampling rate, that is to say, the extra noise contributed by the amplifier may overpower the benefit it brings in.

On the other hand, it is interesting to see whether the resolution is the ultimate restriction of the generation speed. When the resolution is almost infinite, can we get infinite randomness output with the discussed experiment setup? Here, we show that even if the resolution of the ADC can be infinite, the generation speed will not increase infinitely. Given a perfect photon-number resolving detector, the upper bound of the min-entropy is determined by the photon number within the detection time window \cite{MXF:2013:RandomPost}. Suppose that the maximum of the photon number detected per sample is $N$, the total randomness output in the setup, as shown in Fig.~\ref{DEVICE}, is at most $\log_2{(N+1)} \sim \log_2{N}$ (when the photon numbers detected per sample follows a uniform distribution). That is to say, the upper bound of randomness extracted per sample is about $\log_2{N}$. Actually the value of the voltage measured with the ADC is discrete. As the sampling rate increases, the voltage measured by the ADC will finally fix at some certain value and the output randomness will be zero. Therefore, there still exists a limit of the speed of generating random numbers.

Moreover, here we assume that the response time of the photo-detector is small enough to be ignored. In practice, the photo-detector has finite bandwidth, so the detection can be understood as an integral process over a certain time interval, which is the origin of detector response time. If we consider it in our model, we only need to replace $\tau_l$ with $\tau_l+\tau_r$, where $\tau_r$ represents the detector response time satisfying $\tau_s>\tau_r$ and $\tau_c>\tau_r$ \cite{Qi2010}.

\section{Discussion and conclusion}\label{Sec:discussion}
In randomness evaluation, we quantify it with min-entropy. In a sense, we assume that the classical noise is known to the adversary but not malicious. In a recent work, the malicious classical noise scenario is considered by quantifying the randomness with conditional min-entropy  \cite{haw2014maximisation}. Assuming the classical noise $e$ to be also Gaussian, in our min-entropy formula, Eq.~\eqref{Eq:min-entropy}, $\Pr(X)$ should be replaced with a conditional probability $\Pr(X|e)$ to calculate the conditional min-entropy $H_c$. When the adversary is assumed to have full control over the classical noise, the conditional min-entropy minimizing $H_c$ over $e$ should be used in the worst case scenario. On the other hand, when the adversary is restricted to passive eavesdropping, the conditional min-entropy averaging $H_c$ on the distribution of $e$ is used instead. Here we point out that the quantification of randomness in our model is quite similar to the worst-case conditional min-entropy when the variance of $e$ is small enough, which needs more study in the future work.

In conclusion, by reviewing the QRNG scheme and its underlying intuitive physical model, we derive the assumptions and provide a more rigorous models for the QRNG based on vacuum fluctuation. Our result suggest that random number generation speed in such a QRNG is finitely bounded in practice and the device parameters can be optimized to maximize the generation speed.

\section*{Acknowledgments}
The author acknowledges insightful discussions with Z.~Cao, B.~Qi, F.~Xu, and Q.~Zhao. This work was supported by the National Basic Research Program of China Grants No.~2011CBA00300 and No.~2011CBA00301, and the 1000 Youth Fellowship program in China.

%%%%%%%%%%%%%%%%%%%%%%%%%%%%%%%%%%%%%%%%%%%%%%%%%%%%%%%%%%%%%%%%%%%%%%%%%%%%%%%%%%%%%%%%%%%%%%%%%%%%%%%%%%%%%%%%%%%%%%%%
\appendix
%\section{Proof of}

\section{Demonstration of the equivalence of output randomness in two situations: $\tau_s\ll\tau_l$ and $\tau_s=\tau_l'$}
In this appendix, we show the output randomness generated in two situations: $\tau_s\ll\tau_l$ and $\tau_s=\tau_l'$ are the same. First, we need to notice that when the sampling rate is high enough, we can think $\tau_l $  be an integer multiple of $\tau_s$. That is, $\tau_l=n\tau_s$, where $n$ is a positive integer.

\begin{figure}[htbp]
\centering
\resizebox{8cm}{!}{\includegraphics{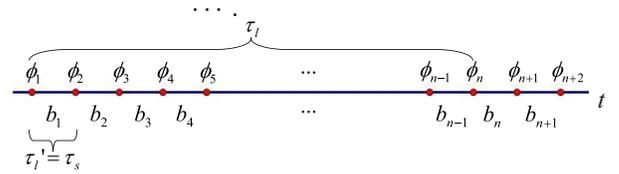}}
\caption{After a exclusive or operation between every two adjacent sequences of random numbers, the new sequences can be expressed as $b_{n+1}-b_1$, $b_{n+2}-b_2$, $\ldots$.}
\label{appendixb}
\end{figure}

Here, denote the phase by $\phi_k$ with label $k$ representing time sequence, then the raw data can be given by
\begin{equation}\label{}
\begin{aligned}
  d_1 &= \phi_{n+1}-\phi_1, \\
  d_2 &= \phi_{n+2}-\phi_2, \\
  &\ldots
  \end{aligned}
\end{equation}
In experiment,  autocorrelation of the random numbers $\{d_k\}$ in the situation $\tau_s\ll\tau_l$ is very high. To reduce the autocorrelation,  ``exclusive OR'' operations can be applied to every two adjacent sequences of random numbers, as shown in Fig.~\ref{appendixb}.
After the exclusive or operation, we can get new sequences of random numbers
\begin{equation}\label{}
\begin{aligned}
  d_1' = d_2 - d_1 &= \phi_{n+2}-\phi_{n+1} - (\phi_2-\phi_1), \\
  d_2' = d_3 - d_2 &= \phi_{n+3}-\phi_{n+2} - (\phi_3-\phi_2), \\
  &\cdots.
\end{aligned}
\end{equation}

Denote $b_k = d_{k+1} - d_{k}$, then we can see from the main context that $b_k$ are independent random numbers. Then, we can express the $d_k'$ by
\begin{equation}\label{}
\begin{aligned}
  d_1' &= b_{n+1} - b_1, \\
  d_2' &= b_{n+2} - b_2, \\
  &\cdots.
\end{aligned}
\end{equation}

Then, we can easily see that the randomness of $\{b_k\}$ is the same to the one of $\{d_k'\}$ in the asymptotic case. While the randomness of $\{b_k\}$ is generated when we have an equivalent interferometer with $\tau_l'=\tau_s$.

%As is discussed above, the interference can be considered to come from two points in the same laser beam with
%time difference $\tau_l$.

%%%%%%%%%%%%%%%%%%%%%%%%%%%%%%%%%%%%%%%%
%\bibliographystyle{ieeetr}
%\bibliographystyle{unsrt}
\bibliographystyle{apsrev4-1}
%%%%%%%%%%%%%%%%%%%%%%%%%%%%%%%%%%%%%%%%

%%%%%%%%%%%%%%%%%%%%%%%%%%%%%%%%%%%%%%%%
% choose a .bib file
\bibliography{BibLaserRandom}
%%%%%%%%%%%%%%%%%%%%%%%%%%%%%%%%%%%%%%%%%%%%%%%%%%%%%%%%%%%%%%%%%%%%%%%%%%%%%%%%%%%%%%%%%%%%%%%%%%%%%%%%%%%%%%%%%%%%%%%%

%\nocite{*}

\end{document}